\documentclass[reprint,superscriptaddress,nofootinbib,amsmath,amssymb,aps,preprintnumbers,prl]{revtex4-2}
\usepackage{graphicx}
\usepackage{color}
\usepackage{dcolumn}
\usepackage{bm}
\begin{document}
\preprint{RIKEN-iTHEMS-Report-22}
\title{Nuclear Many-Body Effect on Particle Emissions Following Muon Capture on $^{28} \mathrm{Si}$ and $^{40} \mathrm{Ca}$}
%
\author{Futoshi Minato}
\email{minato.futoshi@jaea.go.jp}
\affiliation{Nuclear Data Center, Japan Atomic Energy Agency, Tokai, Ibaraki 319-1195 Japan}
\affiliation{RIKEN Nishina Center, Wako, Saitama, 351-0198 Japan}
\author{Tomoya Naito}
\affiliation{RIKEN Interdisciplinary Theoretical and Mathematical Sciences (iTHEMS), Wako, Saitama, 351-0198 Japan}
\affiliation{Department of Physics, Graduate School of Science, The University of Tokyo, Tokyo, 113-0033 Japan}
\author{Osamu Iwamoto}
\affiliation{Nuclear Data Center, Japan Atomic Energy Agency, Tokai, Ibaraki 319-1195 Japan}
\date{\today}
%
\begin{abstract}
  Muon captures on nuclei have provided us with plenty of knowledge of nuclear properties.
  Recently, this reaction attracts attention in electronics, because it is argued that charged particle emissions following muon capture on silicon trigger non-negligible soft errors in memory devices.
  To investigate the particle emissions from a nuclear physics point of view, we develop a new approach using a microscopic model of muon capture and up-to-date particle emission models.
  We paid attention to the muon capture rates, the particle emission spectra, and the multiplicities that have a close interrelation with each other, and found that the nuclear many-body correlation including two-particle two-hole excitations is a key to explaining them simultaneously.
\end{abstract}
\maketitle
\textit{Introduction}---Negative muon capture on nuclei recently attracts high attention due to its various feature.
Muon is firstly captured into outer atomic orbitals generated by the nuclear Coulomb potential and then transits to lower orbitals emitting characteristic muonic X-rays and Auger electrons.
These X-rays are utilized to accurately determine nuclear charge radii~\cite{Fricke1995, Antognini2013,Saito:2022fwi} and are also applied to non-destructive analyses inside vessels~\cite{Shimada-Takaura2021}.
The muon eventually settles in the lowest $s$ orbital and decays via $\mu^{-} \rightarrow e^{-}+\nu_{\mu}+\bar{\nu}_{e}$ or is captured by a nucleus.
The latter case is analogue to electron capture of neutron-deficient nuclei, but can occur even for stable nuclei due to the large muon mass.
Thus, the muon capture is utilized to analyze theoretically calculated nuclear matrix elements of double $\beta$ decay~\cite{Hashim2021}.
%
\par
Another importance of the muon capture is to bring nuclei that captured muon to highly excited states.
Those nuclei deexcite by emitting various particles, and some of which ionize surrounding materials.
Recently, great attention is paid to this process in electronics field because non-negligible soft errors are evoked in memory devices by charged particle and recoiled nucleus after muon capture on silicon~\cite{Serre2012}.
This issue becomes more serious with reducing the scale of memory devices (the so-called die shrink) and operating them at low voltage~\cite{Wang2019}.
Currently, the soft errors in memory devices are studied with Monte-Carlo transport simulations~\cite{Abe2017, Kossov2007} in which, however, the muon capture process is greatly simplified by omitting to solve the nuclear many-body problem.
To date, there is no theoretical framework based on the nuclear structure that describes the muon captures to the particle emissions comprehensively, so developing a more practical model is highly demanded.
%
\par
Particle emissions following the muon capture were extensively studied in the  70--80's (see review article of Ref.~\cite{Measday2001}), and it was discussed that high energy spectra are mainly attributed to instant particle emissions from the preequilibrium state where excitation energy is shared only with a few nucleons in nuclei.
At this stage, it was also pointed out that two-body meson-exchange current (MEC) becomes essential to explain high energy neutron and proton emissions~\cite{Dautry1976, Lifshitz1988}.
In contrast, low energy spectra are attributed to particle evaporation from a compound state where excitation energy is shared with many nucleons in nuclei.
Experimental data of particle emission spectra were explained qualitatively by phenomenological models of preequilibrium and compound states.
However, they have not yet been reproduced accurately enough to be used for a practical application
Moreover, the muon capture process was not discussed carefully from the nuclear structure point of view.
\par
Recent theoretical studies that consider the nuclear many-body systems more appropriately pointed out that correlations resulting from the interaction between nucleons play a significant role in the muon capture~\cite{Lovato2019, Jokiniemi2019, Ciccarelli2020}.
This fact motivated us to revisit the problem of particle emission following muon capture.
To understand the mechanism, we develop a new method that considers the muon capture with a microscopic nuclear model and the particle emission with an up-to-date model of preequilibrium and compound states.
We will demonstrate that many-body correlations resulting from the residual two-body interactions are essential to describe particle emission spectra and multiplicities, as well as muon capture rates.
In particular, the effect of two-particle two-hole ($2p$-$2h$) states, that is to say, the door-way state, is significant for the particle emission spectra.
The target nucleus of this study is $^{28} \mathrm{Si}$, the main material of semiconductors.
In addition, we study $^{40} \mathrm{Ca}$ that have experimental data of emission spectra from low to high energies.
%
%
\par
\textit{Model}---We assume that one-neutron particle one-proton hole ($1p_{\nu}1h_{\pi}$) states are produced in nuclei by the muon capture at first.
We describe this process with the Tamm-Dancoff approximation (TDA) and second TDA (STDA) which is the extension of TDA to $2p$-$2h$ model spaces~\cite{Providencia1965, Minato2016}.
They are extensively applied to study nuclear states at low to high excitation energies including charge-exchange reactions.
Then, we assume that the $1p_{\nu}$-$1h_{\pi}$ state generated by the one-body weak interaction evolves to more complicated multi-particle multi-hole ($mp$-$mh$) states leading to the compound state.
This process, the so-called preequilibrium state, is still one of the challenging subjects in the nuclear physics.
There are mainly two approaches that have been investigated to describe the preequilibrium states, which are semi-classical~\cite{Griffin1966} and quantum models~\cite{FKK} (see also Ref.~\cite{Carlson2014} for the current status). 
The latter model has a good predictive power on several experimental data without phenomenological parameters; however, it is still difficult to describe various kinds of multiple particle emissions in low to high energies comprehensively.
In contrast, the semi-classical model shows a good performance of reproducing various particle emissions, but with some phenomenological ingredients of partial level densities, collision matrices, and so on.
We choose in this work the two-component exciton model~\cite{Koning2004,Iwamoto2016}, one of the semi-classical approaches, because various charged particle emissions are the present scope and the targets in interest are stable nuclei for which the model is well established through the nuclear data evaluations~\cite{Hetrick1997, Shibata2008}.
When nuclei reach the compound state, we describe the particle emission with the Hauser-Feshbach statistical model~\cite{HFSM} that is also applied extensively to study nuclear reactions and particle emission after $\beta$-decay~\cite{Mumpower2018, Minato2021}.
\par
In STDA, excited states of daughter nuclei with spin $J$ are created by operating a phonon creation operator of the vibrational states $Q_{\lambda J}^{\dagger}$ to the target nuclear ground state $\left| 0 \right\rangle$ as
\begin{equation}
  \left|
    \lambda J 
  \right\rangle
  =
  Q_{\lambda J}^{\dagger}
  \left| 0 \right\rangle,
  \qquad
  Q_{\lambda J}
  \left| 0 \right\rangle
  =
  0.
\end{equation}
We calculate the ground state $ \left| 0 \right\rangle$ with the Skyrme-Hartree-Fock (SHF) method~\cite{PhysRevC.5.626} in coordinate space assuming spherical symmetry.
Note that $^{28} \mathrm{Si}$ is considered to form a oblate shape, and the deformation effect and the pairing correlation need to be taken into account practically. 
However, it has been studied that those influences are not striking because of the conservation of closed shell structure for oblate $^{28} \mathrm{Si}$~\cite{Cheoun2017}.
We studied with using two effective forces, which are SGII~\cite{Giai1981} and SkO'~\cite{Reinhard1999}.
Those forces provide a reasonable strength distribution of time-odd channels and the order of low-lying states of $^{28} \mathrm{Al}$ and $^{40} \mathrm{K}$.
Continuum states are discretized by a box size $R=14 \, \mathrm{fm}$.
The STDA phonon operator of the vibrational states is given by
\begin{equation}
  Q_{\lambda J}^{\dagger}
  =
  \sum_{mi} X_{mi}^{\lambda J} 
  \mathcal{O}_{mi}^{J\dagger}
  + 
  \sum_{m\le n, i\le j} \mathcal{X}_{mnij}^{\lambda J} \mathcal{O}_{mnij}^{J\dagger},
  \label{eq:Q}
\end{equation}
where $m$ and $n$ denote particle states, while $i$ and $j$ denote hole states. 
The operators $\mathcal{O}_{mi}^{JM\dagger}$ and $\mathcal{O}_{mnij}^{JM\dagger}$ create $1p$-$1h$ and $2p$-$2h$ states, respectively.
Omitting the second term of Eq.~\eqref{eq:Q} corresponds to the TDA.
The coefficients $X$ and $\mathcal{X}$ in Eq.~\eqref{eq:Q} that effectively reflect the effects of the residual two-body interaction are obtained by solving the TDA and STDA equation~\cite{Providencia1965, Minato2016}.
The model space for $1p$-$1h$ state is set to be $\varepsilon_{m}-\varepsilon_{i} \le 100 \, \mathrm{MeV}$, where $\varepsilon$ is the single-particle energy, and that for $2p$-$2h$ state is restricted to $2n_{q}+l_{q}\le4$ except the $0g$ states, where $n_{q}$ and $l_{q} $ ($ q = \pi $,$ \nu $) are the principle number and orbital angular momentum of a single-particle state. 
\par
Muon capture rates are written as 
\begin{align}
  & \omega \left( E \right)
    =
    \sum_{\lambda, \, J}
    \frac{2 G^2 \nu^2}{1 + \nu / M}
    \notag \\
  & \times
    \left[
    \left|
    \sum_{\nu\pi}
    X_{\nu\pi}^{\lambda J}
    \left\langle
    j_{\nu} l_{\nu}
    \middle\|
    \phi \left( \hat{\mathcal{M}}_J - \hat{\mathcal{L}}_J \right)
    \middle\|
    j_{\pi} l_{\pi}
    \right\rangle
    \right|^2
    \right. 
    \notag \\
  & 
    \left.
    +
    \left|
    \sum_{\nu\pi}
    X_{\nu\pi}^{\lambda J}
    \left\langle
    j_{\nu} l_{\nu}
    \middle\|
    \phi \left( \hat{\mathcal{T}}_J^{\text{el}} - \hat{\mathcal{T}}_J^{\text{mag}} \right)
    \middle\|
    j_{\pi} l_{\pi}
    \right\rangle
    \right|^2
    \right]
    \delta \left( E - E_{\text{x}}^{\text{($ \lambda $, $ J $)}} \right)
    \label{eq:rate}
\end{align}
where $G=1.166\times10^{-11} \, \mathrm{MeV}^{-2}$ is the Fermi coupling constant, $M$ is the mass of the target nucleus, $\nu=m_{\mu}-\Delta M+\varepsilon_{\mu}-E_{\text{x}}^{\text{($ \lambda $, $ J $})}$ is the muon neutrino energy, $\Delta M$ is the mass difference between parent and descendant nuclei taken from AME2020~\cite{Wang2021}, $j_{\nu, \, \pi}$ are the total angular momentum, and $\varepsilon_{\mu}$ is the muon binding energy.
The excitation energy $E_{\text{x}}^{\text{($ \lambda $,$ J $)}}$ with respect to the ground states of the daughter nuclei ($^{28} \mathrm{Al}$ and $^{40} \mathrm{K}$) resulting from the muon capture on the parent nuclei ($^{28} \mathrm{Si}$ and $^{40} \mathrm{Ca}$), respectively, are approximated by $E_{\text{x}}^{\text{($ \lambda $, $ J $)}}=E_{\text{TDA}}^{\text{($ \lambda  $, $ J $)}}-E_{\text{TDA}}^{\text{($ 0 $)}}$, where $E_{\text{TDA}}^{\text{($ \lambda $, $J$)}}$ and  $E_{\text{TDA}}^{\text{($ 0 $)}}$ are the STDA or TDA phonon energies and the lowest energies, respectively.
We consider spin-parity up to $J^{\pi} \le 5^{\pm}$.
The lowest states of TDA and STDA are, respectively, $3^{+}$ and $4^{-}$ for $^{28} \mathrm{Al}$ and $^{40} \mathrm{K}$, which are consistent with the experimental data.
The one-body operators of charge $\hat{\mathcal{M}}_{J}$, longitudinal $\hat{\mathcal{L}}_{J}$, transverse electric $\hat{\mathcal{T}}^{\text{el}}_{J}$, and transverse magnetic $\hat{\mathcal{T}}^{\text{mag}}_{J}$ fields are found in Ref.~\cite{Connel1972}.
For the axial-vector coupling, the free nucleon value is quenched to some extent in nuclei;
however, the exact value is still unknown exactly.
Hence, we study with $g_{A}=-1.26$ (free nucleon) and the quenched value of $g_{A}=-1$.
This quenching is partly explained by the coupling with $2p$-$2h$ states, and we will discuss this point later.
For numerical purpose, $\delta$-function in Eq.~\eqref{eq:rate} is replaced by the Lorentzian function with a width of $1 \, \mathrm{MeV}$.
%
\par
The muon binding energy of $1s_{1/2}$ orbital, $\varepsilon_{\mu}$, and the wave function, $\phi \equiv \phi \left( r \right)$, are calculated by solving the Dirac equation under
the Coulomb potential formed by the atomic nucleus,
$ V_{\text{$ N $-$ \mu $}}^{\text{Coul}} $,
and the $ \left( Z - 1 \right) $ electrons, 
$ V_{\text{$ e $-$ \mu $}}^{\text{Coul}} $,
and the vacuum polarization of the Coulomb potential formed by the atomic nucleus,
$ V_{\text{$ e $-$ \mu $}} $.
Here, $ Z $ denotes the atomic number of the atom.
In this work, the electron density distribution is calculated by the density functional theory~\cite{Hohenberg1964Phys.Rev.136_B864, Kohn1965Phys.Rev.140_A1133} with the local density approximation (LDA),
where the PZ81 LDA correlation functional~\cite{Perdew1981Phys.Rev.B23_5048} is used.
It should be noted that the electron density and the muon wave function are solved simultaneously and self-consistently;
thus, the effect of the muon is also considered in the electron density distribution as well.
The vacuum polarization between the nucleus and the muon 
$ V_{\text{$ N $-$ \mu $}}^{\text{VP}} $
is considered by using the Uehling effective potential~\cite{Uehling1935Phys.Rev.48_55,WayneFullerton1976Phys.Rev.A13_1283}.
Consequently, the muon wave function reads 
\begin{equation}
  \left[
    T 
    +
    V_{\text{$ N $-$ \mu $}}^{\text{Coul}} \left( \bm{r} \right)
    +
    V_{\text{$ N $-$ \mu $}}^{\text{VP}} \left( \bm{r} \right)
    +
    V_{\text{$ e $-$ \mu $}}^{\text{Coul}} \left( \bm{r} \right)
  \right]
  \phi \left( \bm{r} \right)
  =
  \varepsilon
  \phi \left( \bm{r} \right),
\end{equation}
where $ T $ is the Dirac kinetic operator.
For more detail, see Supplemental Material.
\par
Assuming that the $1p_{\nu}$-$1h_{\pi}$ state generated by the one-body operators of the muon capture as an initial state, we carry out the two-component exciton model calculation.
The master equation of the two-component exciton model is~\cite{Iwamoto2016}
\begin{align}
  \frac{dP \left( p_{\pi}, p_{\nu}, t \right)}{dt}
  & = P \left( p_{\pi}-1,p_{\nu},t \right)
    \lambda_{\pi+} \left( p_{\pi}-1,p_{\nu} \right)
    \notag \\
  & + 
    P \left( p_{\pi},p_{\nu}-1,t \right)
    \lambda_{\nu+} \left( p_{\pi},p_{\nu}-1 \right)
    \notag \\
  & + 
    P \left( p_{\pi}-1,p_{\nu}+1,t \right)
    \lambda_{\nu\pi} \left( p_{\pi}-1,p_{\nu}+1 \right)
    \notag \\
  & + 
    P \left( p_{\pi}+1,p_{\nu}-1,t \right)
    \lambda_{\pi\nu} \left( p_{\pi}+1,p_{\nu}-1 \right)
    \notag \\
  & - 
    P \left( p_{\pi},p_{\nu},t \right)
    \left[
    \lambda \left( p_{\pi},p_{\nu} \right)
    +
    W \left( p_{\pi},p_{\nu} \right)
    \right],
    \label{eq:two-exciton}
\end{align}
where $\lambda \left( p_{\pi},p_{\nu} \right)
=\sum_x \lambda_x \left( p_{\pi},p_{\nu}\right) $ is the total transition rate,
$\lambda_{q+}$ is the creation rate of a particle-hole pair, $\lambda_{\pi\nu}$ is the exchange rate of proton and neutron particle-hole pairs, and $P \left( p_{\pi},p_{\nu},t \right)$ is the occupation probability of the exciton state having proton particle number $p_{\pi}$ and neutron particle number $n_{\nu}$ at time $t$.
$W \left( p_{\pi},p_{\nu} \right) = \sum_b W_{b} \left( p_{\pi},p_{\nu} \right)$ is the total particle emission rate, and $W_{b}$ is the particle emission rate for the particle $b$ calculated with the inverse reaction cross section and the partial level density for residual nuclei. Here, notation of hole state is omitted for simplicity.
%
By solving Eq.~\eqref{eq:two-exciton} with initial condition of $1p_{\nu}$-$1h_{\pi}$, the probability for emitting particle $b$ can be calculated by
\begin{equation}
  \sigma \left( E \right)
  =
  R \left( E \right)
  \sum_{p_{\pi}, p_{\nu}}
  Q \left(p_{\pi},p_{\nu}\right)
  W_{b} \left(p_{\pi},p_{\nu} \right)
\end{equation}
where $Q$ is the cumulative occupation probability defined as
$Q \left( p_{\pi},p_{\nu} \right)
=
\int_{t_{0}}^{t_{1}} P \left( p_{\pi},p_{\nu}\right) \, dt$,
$R$ the normalized capture rate as
$R \left( E \right) = \omega \left( E \right) / \int \omega \left( E' \right) \, dE'$ 
and $ W_{b,p} \left( p_{\pi}, p_{\nu} \right) $
is the particle emission rate from the exciton state
$ \left( p_{\pi}, p_{\nu} \right) $. 
As the number of exciton is greater than $12$, the calculation of preequilibrium state terminates and that of compound state initiates.
We use a standard parameter set for preequilibrium and compound states that is globally used in evaluating nuclear data (see Ref.~\cite{Iwamoto2016} for more detail).
Only for the proton and neutron single-particle state densities $g$, we adjust to $g_{\pi}=Z/19$ and $g_{\nu}=N/19$ from the standard value of the Fermi gas, $g_{\pi}=Z/15$ and $g_{\nu}=N/15$.
This adjustment is reasonable considering the semi-magic structure of $^{28} \mathrm{Si}$ and the magic structure of $^{40} \mathrm{Ca}$ (see Supplemental Material).
\par
\textit{Result and discussion}---Table~\ref{tab:rate} lists the calculated muon capture rates of $^{28} \mathrm{Si}$ and $^{40} \mathrm{Ca}$.
In addition to TDA and STDA, we also show the result of ``FREE'' that is obtained by assuming that nucleons move independently in the nuclear potential.
The experimental data for natural silicon and calcium are also listed.
Note that the natural abundances of $^{28}$Si and $^{40}$Ca are about $92$\% and $97$\%, respectively, and their muon capture rates are expected to be close to those for the natural elements.
The range of calculated muon capture rates is estimated with the axial-vector coupling of $g_{A}=-1$ and $-1.26$, which correspond to the lower and upper values of the calculated muon capture rates, respectively.
We find that the muon capture rates of FREE overestimate experimental data of $^{28} \mathrm{Si}$ and $^{40} \mathrm{Ca}$ both for SGII and SkO', while TDA and STDA reproduces reasonably well within the uncertainties of $g_{A}$.
One may notice that the lower value of the capture rate, that is the result of $g_{A}=-1.0$, is favorable for TDA, while the result of weakly quenched axial-vector coupling is supportive for STDA.
This result is consistent to a picture that the quenching of the axial-vector coupling is partly explained by the coupling with $2p$-$2h$ states~\cite{Ichimura2006}.
\begin{table}[b]
  \centering
  \caption{Calculated muon capture rates for $^{28} \mathrm{Si}$ and $^{40} \mathrm{Ca}$ (in the unit of $10^{6} \, \mathrm{s}^{-1}$). The lower and upper values of model calculations are obtained by using $g_{A}=-1$ and $-1.26$, respectively. The experimental data for natural silicon and calcium are also listed~\cite{Suzuki1987}. The numbers in parentheses are the uncertainties of the corresponding last digits.}
  \label{tab:rate}
  \begin{ruledtabular}
    \begin{tabular}{ll|ccc|l}
      \multicolumn{1}{c}{Nucl.} & \multicolumn{1}{c|}{Force} & \multicolumn{1}{c}{FREE} & \multicolumn{1}{c}{TDA} & \multicolumn{1}{c|}{STDA} & \multicolumn{1}{c}{Expt.} \\
      \hline
      $^{28} \mathrm{Si}$ & SGII & $1.02$--$1.46$ & $0.87$--$1.26$ & $0.81$--$1.18$ &$0.8712 \left( 18 \right)$ \\
                                & SkO' & $1.04$--$1.49$ & $0.87$--$1.26$ & $0.72$--$1.04$\\
      $^{40} \mathrm{Ca}$ & SGII & $3.12$--$4.35$ & $2.54$--$3.58$ & $2.44$--$3.32$ & $2.557 \left( 14 \right)$ \\
                                & SkO' & $2.58$--$3.63$ & $2.07$--$2.95$ & $1.90$--$2.58$
    \end{tabular}
  \end{ruledtabular}
\end{table}
\par
\begin{figure}
  \centering
  \includegraphics[width=1\linewidth]{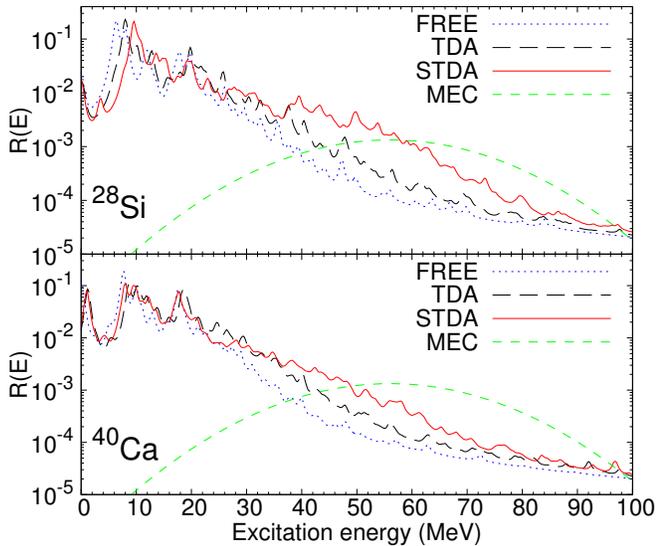}
  \caption{Normalized capture rates $R \left( E \right)$ for $^{28} \mathrm{Si}$ (top) and $^{40} \mathrm{Ca}$ (bottom) obtained by FREE, TDA, and STDA. The results of SkO' force with $g_{A}=-1.0$ are shown. The horizontal axis represents the excitation energy of $^{28}\mathrm{Al}$ and $^{40}\mathrm{K}$. Contributions of MEC approximated by a Gaussian form are also shown by the dashed line (see text).}
  \label{fig:exc}
\end{figure}

The agreement with experimental data is related to the many-body correlations resulting from the residual two-body interaction.
To explain it, we show in Fig.~\ref{fig:exc} the normalized capture rates $R \left( E \right)$ of $^{28}\mathrm{Al}$ and $^{40}\mathrm{K}$ calculated by SkO' (see Supplemental Material for that by SGII).
The functions of $R \left( E \right)$ for FREE, TDA, and STDA are similar to each other in excitation energies less than $20$ MeV, where are the major part of the capture rates having more than $1 \, \%$.
However, those of TDA and STDA distribute in higher excitation energies than FREE by a few MeV, and moreover, significant enhancements above $E=25 \, \mathrm{MeV}$ are observed.
This is because the residual two-body interaction works repulsively for most of the $J^{\pi}$ channels.
Such a feature substantially decreases the momentum of outgoing muon neutrino, reducing the muon capture rate through the factor $\nu^2$ in Eq.~\eqref{eq:rate}.
\par
STDA gives additional enhancements above $E=30 \, \mathrm{MeV}$ as compared with TDA as seen in Fig.~\ref{fig:exc}.
This is because the $1p$-$1h$ states couple with the $2p$-$2h$ ones, and some of them at high energies receive substantial strengths from those at low energies.
The enhancement of transition probabilities at high energies induces the further reduction of muon capture rates as found in Table~\ref{tab:rate}, and the weak quenching of the axial-vector coupling shows a favorable agreement with the experimental data accordingly.
Although the capture rates above $30\,\mathrm{MeV}$ are about $1 \, \%$ at most and the enhancement due to the $2p$-$2h$ states is a little, they affect particle emission spectra appreciably as discussed next.
In general, the effect of the residual interaction becomes small with increasing energy, and the capture rates of FREE, TDA, and STDA get closer to each other.
\par
\begin{figure}
  \centering
  \includegraphics[width=0.89\linewidth]{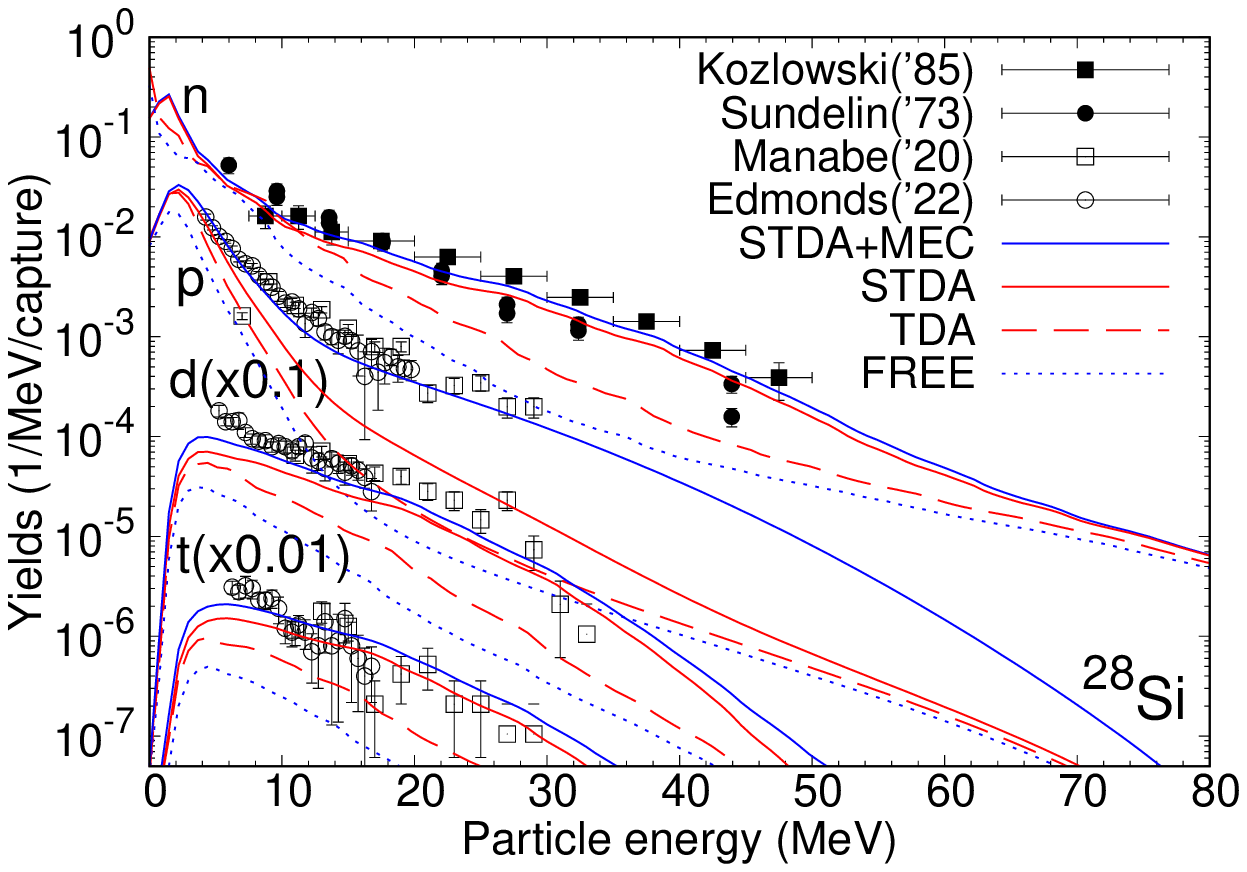}
  \includegraphics[width=0.89\linewidth]{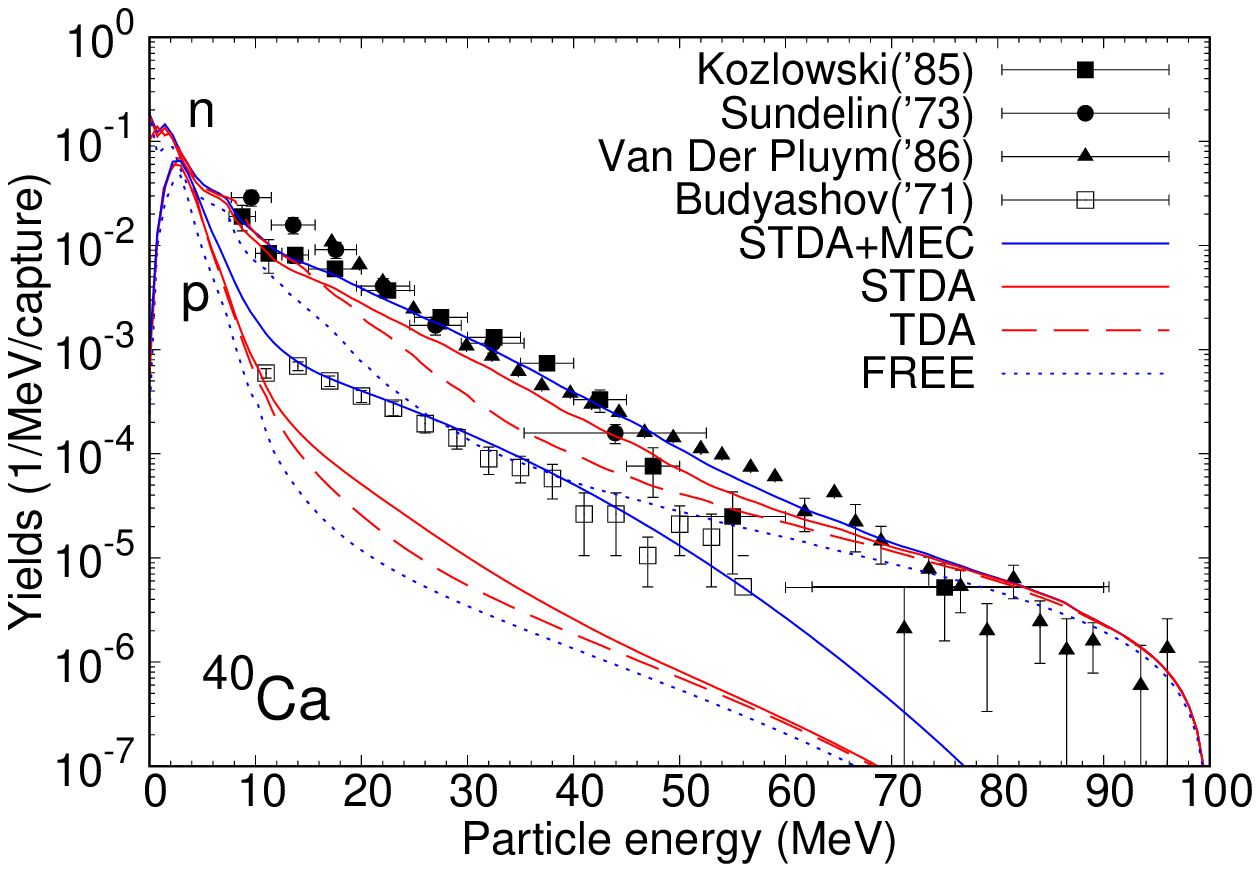}
  \caption{Particle yields after the muon capture on $^{28}\mathrm{Si}$ (top) and $^{40}\mathrm{Ca}$ (bottom).
    The result of SkO' for $g_{A}=-1$ is shown. Experimental data of neutron (filled symbols) and proton (open symbols) are taken from Refs.~\cite{Sundelin1973, Kozlowski1985, VanDerPluym1986} and \cite{ Budyashov1971, Edmonds2022, Manabe2020}, respectively.
    Note that the unit is not given in the original paper of Budyashov~\cite{Budyashov1971}, so that we normalized the second point from the low energy to STDA+MEC.}
  \label{fig:spec}
\end{figure}

Figure~\ref{fig:spec} shows the particle emission spectra of muon capture on $^{28}\mathrm{Si}$ and $^{40}\mathrm{Ca}$.
The results of SkO' with the axial-vector coupling $g_{A}=-1$ are illustrated together with the experimental data. 
The peaks formed around $E=2$--$4\,\mathrm{MeV}$ are resulting from the particle emission from the compound state, while high energy tails of spectra are due to the particle emission from the preequilibrium state.
The result of FREE underestimates largely the experimental data.
This shortcoming is improved by TDA because of the enhancement of the capture rates at high energies as seen in Fig.~\ref{fig:exc}.
STDA further raises the calculated spectra and the results get more closer to experimental data.
We would like to stress here that the effect of the coupling of $2p$-$2h$ states was only less than $1 \, \%$ for muon capture rates as seen in Fig.~\ref{fig:exc};
however, its influence on particle emission spectra (Fig.~\ref{fig:spec}) is non-negligible.
\begin{table*}[bth]
  \centering
  \caption{Calculated multiplicities per $10^{3}$ muon captures for $^{28} \mathrm{Si}$ and $^{40} \mathrm{Ca}$, compared with the experimental data for the natural silicon and calcium. The results of SGII and SkO' forces with $g_{A}=-1$ are shown. Experimental data of neutron are taken from Ref.~\cite{Macdonald1965}, while those of charged particles are from Ref.~\cite{Edmonds2022}. The numbers in parentheses are the uncertainties of the corresponding last digits.}
  \label{tab:multiplicity}
  \begin{ruledtabular}
    \begin{tabular}{lcc|dddddddd|c}
      \multicolumn{1}{c}{Nucl.} & \multicolumn{1}{c}{Particle} & \multicolumn{1}{c|}{Energy} & \multicolumn{2}{c}{FREE} & \multicolumn{2}{c}{TDA} & \multicolumn{2}{c}{STDA} & \multicolumn{2}{c|}{STDA+MEC} & \multicolumn{1}{c}{Expt.} \\
                                & & \multicolumn{1}{c|}{Range ($ \mathrm{MeV} $)} & \multicolumn{1}{c}{SGII} & \multicolumn{1}{c}{SkO'} & \multicolumn{1}{c}{SGII} & \multicolumn{1}{c}{SkO'} & \multicolumn{1}{c}{SGII} & \multicolumn{1}{c}{SkO'} & \multicolumn{1}{c}{SGII} & \multicolumn{1}{c|}{SkO'} & \\
      \hline
      $^{28} \mathrm{Si}$ 
                                & $n$ & Entire & 639 & 573 & 631 & 812 & 676 & 938 & 729 & 983 & $864 \left( 72 \right)$ \\
                                & $p$ & $4$--$20$ & 14.9 & 13.5 & 23.9 & 23.7 & 31.8 & 41.4 & 54.4 & 63.5 & $52.46 \left( 192 \right)$\\
                                & $d$ & $5$--$17$ & 1.99 & 2.01 & 3.98 & 4.13 & 5.89 & 7.13 & 7.96 & 9.15 & $9.80 \left( 46 \right)$\\
                                & $t$ & $6$--$17$ & 0.303 & 0.309 & 0.714 & 0.720 & 1.18 & 1.54 & 1.63 & 1.97 & $1.70 \left( 13 \right)$\\
                                & $\alpha$ & $15$--$20$ & 0.288 & 0.285 & 0.598 & 0.607 & 0.944 & 1.14 & 1.28 & 1.47 & $0.57 \left( 10 \right)$\\
      \hline
      $^{40} \mathrm{Ca}$ & $n$ & Entire & 352 & 473 & 583 & 659 & 559 & 631 & 613 & 681 & $764 \left( 32 \right)$ \\
    \end{tabular}
  \end{ruledtabular}
\end{table*}
\par
In Fig.~\ref{fig:spec}, 
STDA still underestimates the proton spectra largely.
The one-body operators of muon capture in Eq.~\eqref{eq:rate} create only $1p_{\nu}$-$1h_{\pi}$ state.
To enhance the proton emission spectra from this initial state, neutron must give its energy instantly to proton during the preequilibrium state;
however, it is difficult to simulate this energy transfer enough to reproduce experimental data.
In this respect, Lifshitz and Singer have discussed that MEC plays an important role to explain the experimental proton spectra~\cite{Lifshitz1988}.
We also consider the effect phenomenologically in the present framework as follows.
The capture rate due to MEC is almost universal for nuclei according to Lifshitz and Singer, and we approximate it by a Gaussian function, 
where the mean excitation energy and the width of which are given as $56$ and $15\,\mathrm{MeV}$, respectively.
The contribution of MEC to the capture rate is set to be $ 5 \, \%$ as estimated by Lifshitz and Singer~\cite{Lifshitz1988}, while that of STDA is reduced by $ 95 \, \%$.
The capture rate of MEC is illustrated in Fig.~\ref{fig:exc}.
Since MEC is two-body current and has a form of two-body isospin operator $ \left[ \tau_{1} \otimes \tau_{2} \right]^{1,-1}$, 
we assume that two configurations that are $1p_{\pi}$-$2h_{\pi}$-$1p_{\nu}$-$0h_{\nu}$ and $0p_{\pi}$-$1h_{\pi}$-$2p_{\nu}$-$1h_{\nu}$ states are created with the equal probability after the muon capture, and set them as the initial preequilibrium state.
The result of considering MEC (STDA+MEC) is shown in Fig.~\ref{fig:spec}.
Particle emission spectra at high energies are enhanced further, in particular, a remarkable improvement is obtained for the proton spectra.
\par
Table~\ref{tab:multiplicity} lists the calculated and experimental data of multiplicities of emitted particles for $^{28} \mathrm{Si}$ and $^{40} \mathrm{Ca}$.
Here, we used $g_{A}=-1$ in the calculation; however, we confirmed that the difference from $g_{A}=-1.26$ is less than $3 \, \%$.
The results of FREE underestimate the experimental data, while we obtain improvements with increasing the many-body correlations of $1p$-$1h$ mixture (TDA), and the coupling with $2p$-$2h$ states (STDA).
The effect of MEC is comparable to TDA and STDA, making the calculated results even closer to the experimental data.
On the contrary, the calculated $\alpha$ particle multiplicity of $^{28} \mathrm{Si}$ and neutron multiplicity of $^{40} \mathrm{Ca}$ deviate from the experimental data.
The measured energy range of $\alpha$ particle multiplicity for $^{28} \mathrm{Si}$ is limited only to $15$--$20 \, \mathrm{MeV}$ and the available experimental data of neutron multiplicity for $^{40} \mathrm{Ca}$ is only one, which was measured more than $50$ years ago~\cite{Macdonald1965}.
The multiplicity strongly depends on the distribution of capture rates that modulates particle emissions from the preequilibrium state which emits only a few particles and the compound state which emits multiple particles.
For further understanding of particle emissions and validations of the nuclear model, more experimental studies that cover the spectra from low to high energies are required.
The calculated multiplicities of charged particles for entire energy range are provided in Supplemental Material.
%
\par
\textit{Summary}---We demonstrated that the many-body correlation resulting from the residual two-body interaction operating between nucleons played an important role in the muon capture rates of $^{28}\mathrm{Si}$ and $^{40}\mathrm{Ca}$.
In particular, the coupling with $2p$-$2h$ states was essential to describe the particle emission spectra and the multiplicities.
In other words, this indicates that particle emissions following muon captures have information on the nuclear structure at high energies.
The present study showed that the combination of the microscopic approach of muon capture and the two-component exciton model of particle emission is an effective tool to describe particle emission following the muon captures, giving the nuclear structure information additionally.
Only the effect of MEC was considered within a simple manner so that it is demanded to take into account in STDA in future.
The present outcomes are expected to contribute to the development of Monte-Carlo transport simulations of muon captures~\cite{Abe2017, Kossov2007} and further understanding of the nuclear structure.
\par
The $\alpha$ particle multiplicity of $^{28}$Si and neutron multiplicity of $^{40}\mathrm{Ca}$ are the remaining questions of this work.
Currently, three facilities providing negative muon beams (TRIUMF, RAL, and J-PARC) are running in the world, and new experiments are planned there.
In addition, a new negative muon facility is going to launch in RCNP at Osaka Univ.
Those activities will increase information on the interaction between muon and nuclei, and help us not only to verify the present framework but also to develop theoretical models.
\par
\begin{acknowledgments}
  The authors thanks Prof.~S.~Kawase, Dr.~A.~Abe, and Prof.~G.~Col\`{o} for fruitful discussion. FM thanks Dr.~T.~Fukui for supporting this work.
  TN thanks Dr.~S.~A.~Sato for discussion on the code for electron wave functions in the density functional theory.
  TN acknowledges the financial support from the RIKEN Special Postdoctoral Researchers Program.
  The numerical calculations were partly performed on cluster computers at the RIKEN iTHEMS program.
\end{acknowledgments}
%
%
\bibliography{muon}
\end{document}


\title{Supplement of Many-Body Effects on Particle Emission Following Muon Capture on $^{28} \mathrm{Si}$ and $^{40} \mathrm{Ca}$}
\author{Futoshi Minato, Tomoya Naito, and Osamu Iwamoto}
\date{}
\maketitle
\renewcommand{\thefigure}{S\arabic{figure}}
\renewcommand{\thetable}{S\Roman{table}}
\section{Coulomb potentials}
\par
The muon binding energy of $1s_{1/2}$ orbital, $\varepsilon_{\mu}$, and the wave function, $\phi \equiv \phi \left( r \right)$, are calculated by solving the Dirac equation under the Coulomb potential formed by the atomic nucleus and the $ \left( Z - 1 \right) $ electrons.
where $ Z $ denotes the atomic number of the atom.
The charge density distribution of the atomic nucleus is considered in the Coulomb potential,
which reads
\begin{align}
  V_{\text{$ N $-$ \mu $}}^{\text{Coul}} \left( \bm{r} \right)
  &  =
    - 4 \pi
    \int_r^{\infty}
    \frac{1}{r'^2}
    \int_0^{r'}
    \overline{\rho}_{\text{ch}} \left( r' \right)
    r' \, dr'
    \notag \\
  & = 
    - \frac{4 \pi}{r}
    \int_0^r
    \overline{\rho}_{\text{ch}} \left( r' \right)
    r'^2 \, dr'
    -
    4 \pi
    \int_r^{\infty}
    \overline{\rho}_{\text{ch}} \left( r' \right)
    r' \, dr',
\end{align}
where $ \overline{\rho}_{\text{ch}} $ is the spherical-averaged charge density distribution of the atomic nucleus.
The Coulomb potential formed by the $ \left( Z - 1 \right) $ electrons reads
\begin{align}
  V_{\text{$ e $-$ \mu $}}^{\text{Coul}} \left( \bm{r} \right)
  & =
    4 \pi
    \int_r^{\infty}
    \frac{1}{r'^2}
    \int_0^{r'}
    \rho_e \left( r' \right)
    r' \, dr' 
    \notag \\
  & = 
    \frac{4 \pi}{r}
    \int_0^r
    \rho_e \left( r' \right)
    r'^2 \, dr'
    +
    4 \pi
    \int_r^{\infty}
    \rho_e \left( r' \right)
    r' \, dr',
\end{align}
where $ \rho_e $ is the electron density distribution.
%
%
%
%
%
%
%
\section{Single-particle state density}
\begin{figure}[b]
  \centering
  \includegraphics[width=0.4\linewidth]{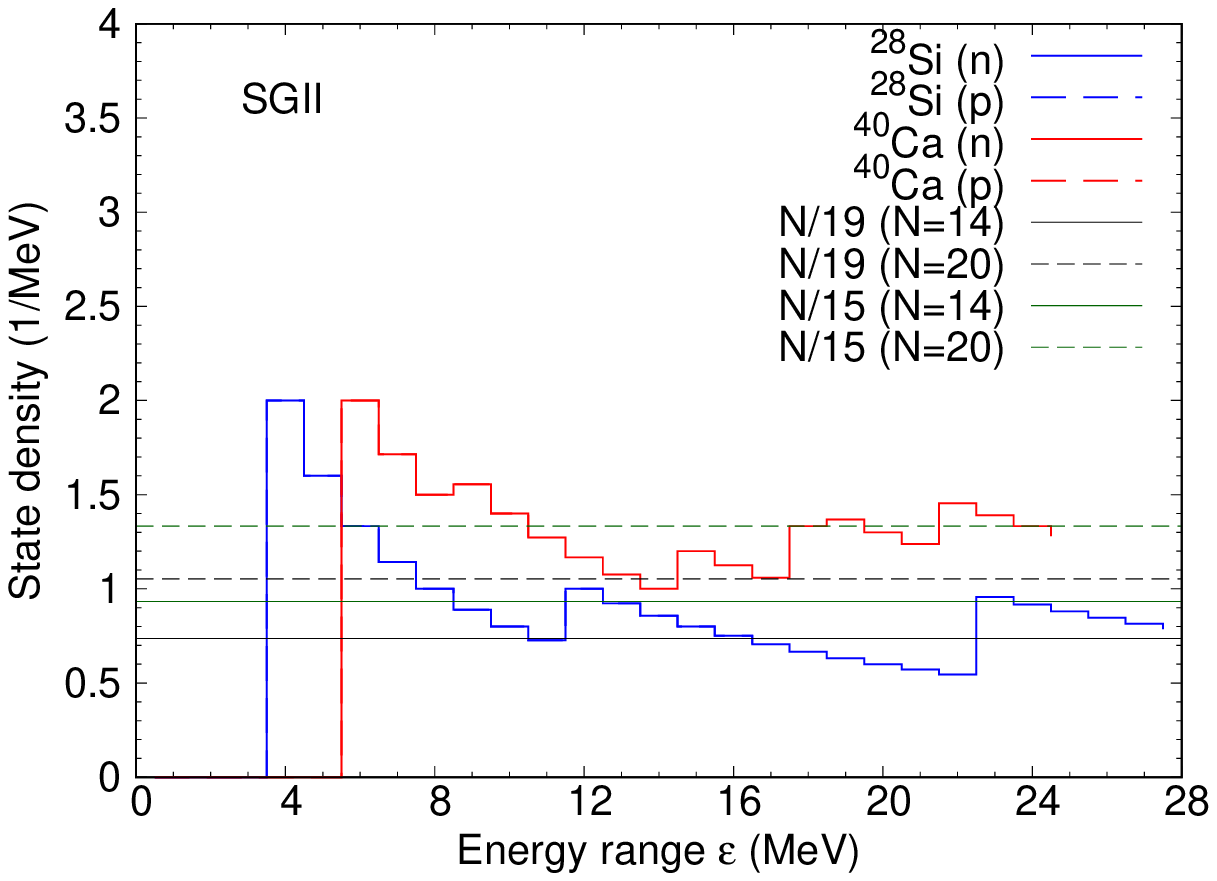}
  \includegraphics[width=0.4\linewidth]{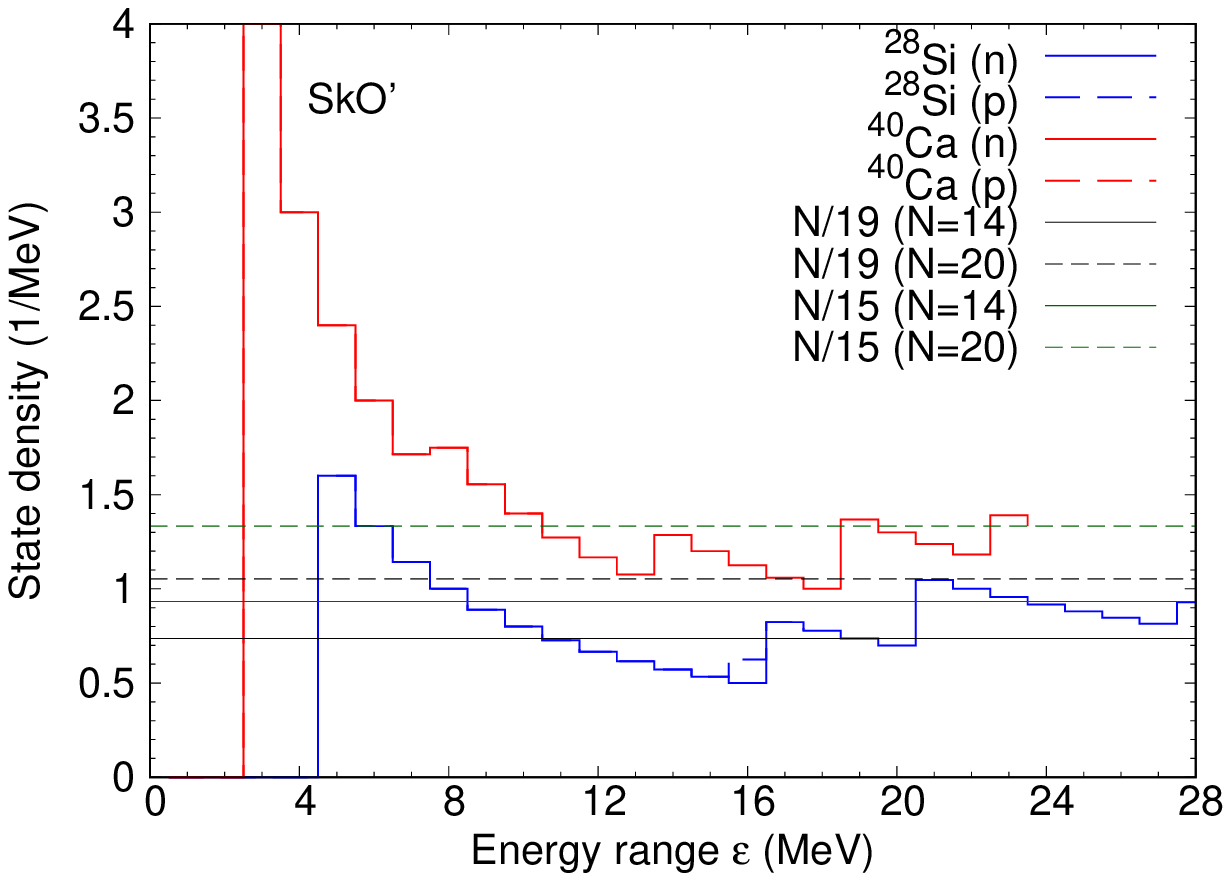}
  \caption{Single-particle state densities of SGII (left) and SkO' (right) as a function of energy range $\varepsilon$. Note that those of neutron and proton are almost overlapped.}
  \label{fig:ssd}
\end{figure}
\par
To calculate the two-component exciton model numerically, one needs the single-particle state densities of neutrons ($g_{\nu}$) and protons ($g_{\pi}$).
These values are usually estimated from that of the Fermi-gas model at the Fermi energies, which are given by $g_{\nu}\simeq N/15$ and $g_{\pi}\simeq Z/15$~\cite{BohrMottelson}.
However, we used $g_{\nu}\simeq N/19$ and $g_{\pi}\simeq Z/19$ rather than those of the Fermi-gas model in this letter, considering the closed-shell structure of $^{28} \mathrm{Si}$ and $^{40} \mathrm{Ca}$.
\par
To validate this choice, we estimated the single-particle state density with the Skyrme-Hartree-Fock (SHF), which is defined as
\begin{equation}
  w \left( \varepsilon \right)
  =
  \frac{\int_{\varepsilon_{f} - \varepsilon/2}^{\varepsilon_{f} + \varepsilon/2}
    \sum_{i}
    n_i 
    \delta \left( \varepsilon_{i} - \varepsilon' \right)
    \,
    d \varepsilon'}{\varepsilon},
\end{equation}
where $\varepsilon_{f}$ the Fermi energy, $\epsilon_{i}$ the single-particle energy of bound state calculated by SHF, and $n_{i}=2j_{i}+1$ the degeneracy.
The Fermi energy is computed by averaging the single-particle energies of the last occupied and first unoccupied levels.
The results of the single-particle state density with an energy bin $\Delta \varepsilon=0.5 \, \mathrm{MeV}$ is shown in Fig.~\ref{fig:ssd}, where those for $N/15$ and $N/19$ are also drawn.
Note that the results of neutron and proton are almost overlapped.
Since the level structure is discrete, the state density is zero when $\varepsilon$ is close to $0$.
Increments observed in the state density occur when a single-particle state is involved within $\varepsilon_{f}\pm\varepsilon/2$.
Actually, it is vague what to extent we should include energy range of $\varepsilon$ to determine an appropriate single-particle state density.
However, we may exclude small $\varepsilon$ where only a few levels are involved and large $\varepsilon$ where is too far from $\varepsilon_{f}$. 
It may be reasonable to select $\varepsilon > 8 \, \mathrm{MeV}$ where the state density settles down to some extent and $\varepsilon < 18 \, \mathrm{MeV}$ since $\varepsilon_{f}$ for proton is about $9 \, \mathrm{MeV}$.
We can see that the line of $N/19$ is closer to the calculated state densities in the wide range of $10\le\varepsilon\le18 \, \mathrm{MeV}$ than $N/15$.
We have also calculated the particle emission spectra with $N/15$. However, the result largely underestimated the experimental data because a large single-particle state density prompts transitions to the compound state and high energy spectra attributed from the preequilibrium state are hindered.
\section{Muon Capture Rate and Particle Emission Spectra of SGII}
\begin{figure*}[bht]
  \centering
  \includegraphics[width=0.49\linewidth]{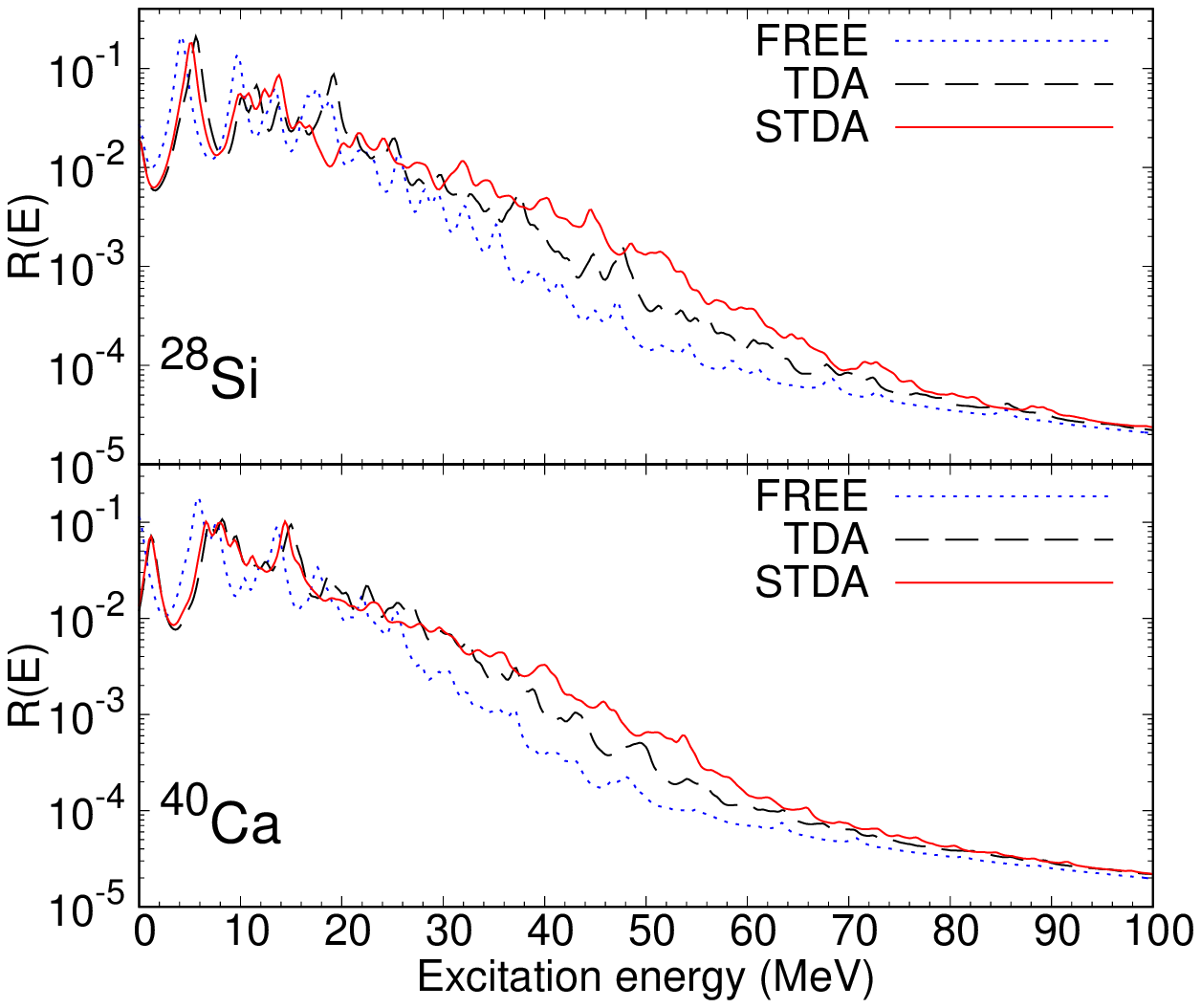}
  \caption{Normalized capture rate $R \left( E \right)$ for $^{28} \mathrm{Si}$ (top) and $^{40} \mathrm{Ca}$ (bottom) obtained by FREE, TDA, and STDA. The result of SGII with $g_{A}=-1$ is shown.}
  \label{fig:exc}
\end{figure*}

\begin{figure*}[bht]
  \centering
  \includegraphics[width=0.49\linewidth]{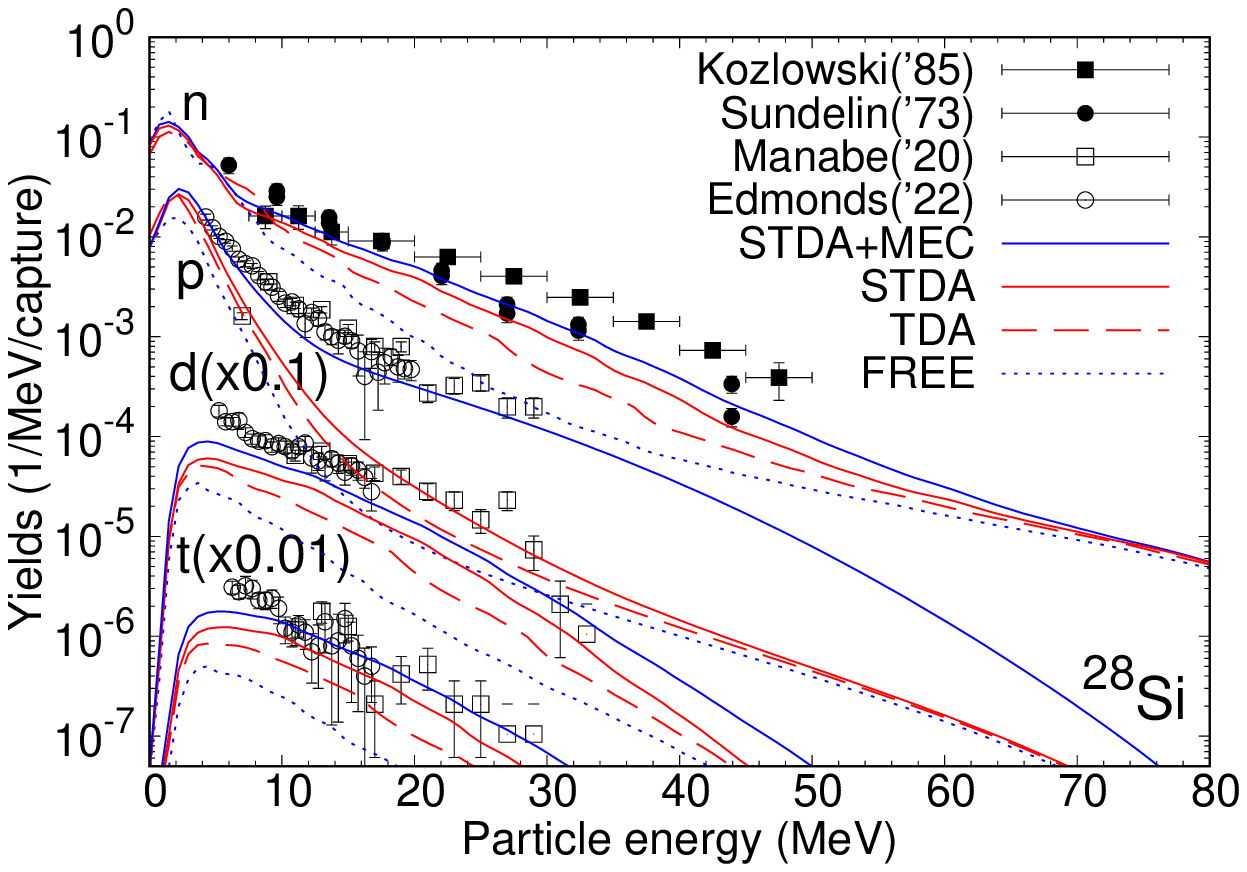}
  \includegraphics[width=0.49\linewidth]{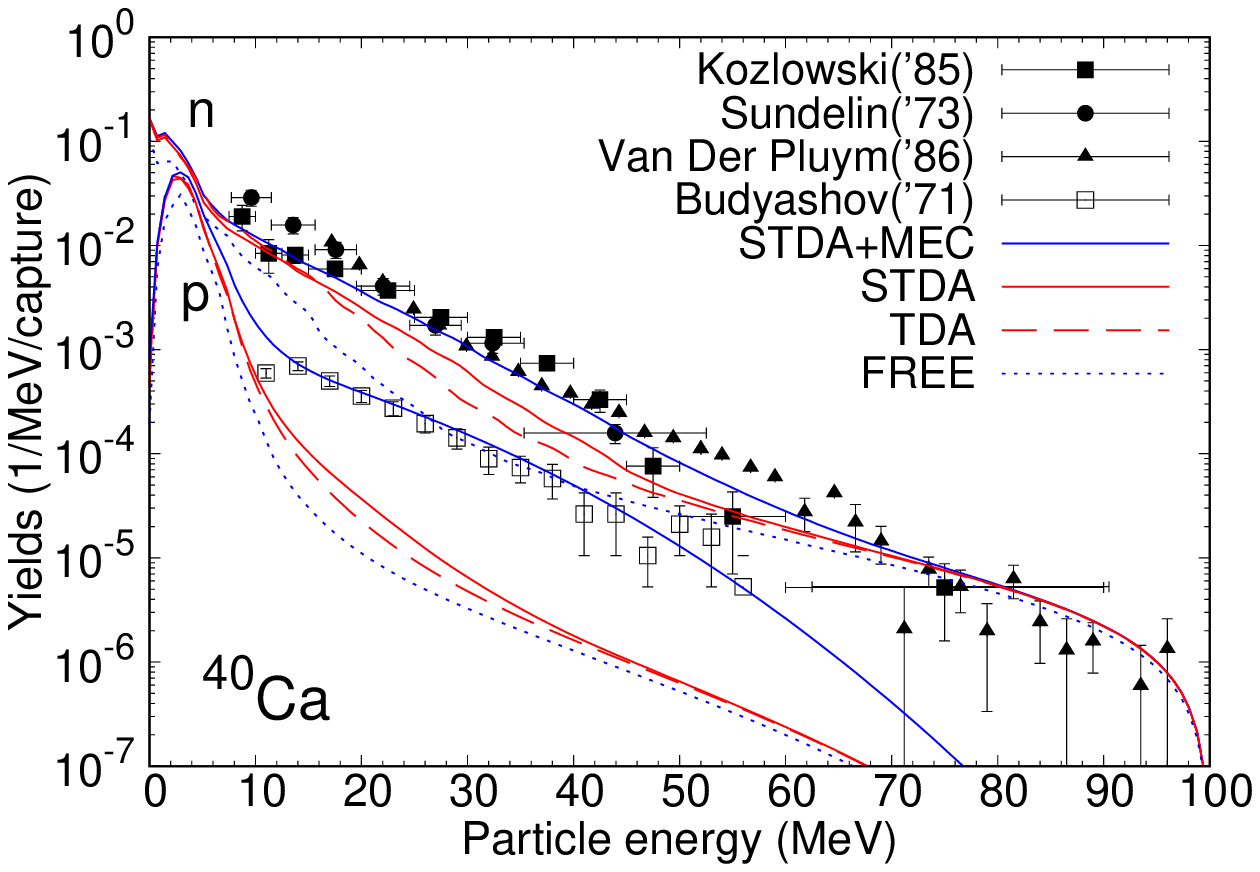}
  \caption{Particle yields after the muon capture on $^{28}\mathrm{Si}$ (left) and $^{40}\mathrm{Ca}$ (right).
    The result of SGII with $g_{A}=-1$ is shown. Experimental data of neutron (filled symbols) and proton (open symbols) are taken from Refs.~\cite{Sundelin1973, Kozlowski1985, VanDerPluym1986} and \cite{Budyashov1971, Edmonds2022, Manabe2020}, respectively.
    Note that unit is not given in the original paper of Budyashov~\cite{Budyashov1971}, so that we normalized the second point from the low energy to STDA+MEC.}
  \label{fig:spec}
\end{figure*}

In the main article, we give only the muon capture rates and spectra calculated by SkO'.
Here, we show the result of SGII.
Figure~\ref{fig:exc} depicts the capture rate.
The result is qualitatively same as SkO'; 
however, the enhancement in energy range $30\le E \le 60 \, \mathrm{MeV}$ due to the coupling with $2p$-$2h$ states of STDA is moderate compared to SkO'.
Figure~\ref{fig:spec} illustrate the paticle emission spectra.
Again, the overall feature is qualitatively same as SkO'.
However, as compared to SkO', SGII slightly underestimates particle emission spectra. 
Consequently, the multiplicities of SGII tend to be smaller than SkO' as seen in Table II of the main article and Table~\ref{tab:multiplicity} below.
\section{Multiplicities of charged particles for entire energy region}
\par
In the main article, the multiplicities of charged particles of $^{28}\mathrm{Si}$ are estimated in the limited energy range.
Here, we show the multiplicities of them for entire energy range in Table~\ref{tab:multiplicity}, where $g_{A}=-1$ is used in the calculation.
In addition to $^{28} \mathrm{Si}$, the result of $^{40} \mathrm{Ca}$ is also listed.
We confirmed that the difference from $g_{A}=-1.26$ is less than $3\,\% $.
Compared to Table II of the main article, estimated in the limited energy range, the multiplicities in the entire energy region are considerably large.
Namely, the measured energy ranges are very narrow to discuss the influences, and therefore further experimental investigates are needed.
\begin{table*}[hbt]
  \centering
  \caption{Calculated and experimental multiplicities per $10^{3}$ muon captures for $^{28} \mathrm{Si}$ and $^{40} \mathrm{Ca}$ in case of the entire energy range. The results of SGII and SkO' forces with the axial-vector coupling $g_{A}=-1$ are shown.}
  \label{tab:multiplicity}
  \begin{tabular}{lc|D{.}{.}{4}D{.}{.}{4}D{.}{.}{4}D{.}{.}{4}D{.}{.}{4}D{.}{.}{4}D{.}{.}{4}D{.}{.}{4}}
    \hline \hline
    \multicolumn{1}{c}{Nucl.} & \multicolumn{1}{c|}{Emitted} & \multicolumn{2}{c}{FREE} & \multicolumn{2}{c}{TDA} & \multicolumn{2}{c}{STDA} & \multicolumn{2}{c}{STDA+MEC} \\
                              & \multicolumn{1}{c|}{particle} & \multicolumn{1}{c}{SGII} & \multicolumn{1}{c}{SkO'} & \multicolumn{1}{c}{SGII} & \multicolumn{1}{c}{SkO'} & \multicolumn{1}{c}{SGII} & \multicolumn{1}{c}{SkO'} & \multicolumn{1}{c}{SGII} & \multicolumn{1}{c}{SkO'} \\
    \hline 
    $^{28} \mathrm{Si}$ 
                              & $p$      & 60.7 & 61.5 & 94.2 & 98.9 & 101 & 119 & 128 & 144 \\
                              & $d$      & 2.51 & 2.49 & 4.68 & 4.91 & 6.71 & 8.66 & 9.32 & 11.2 \\
                              & $t$      & 0.442 & 0.448 & 0.940 & 0.979 & 1.52 & 2.07 & 2.16 & 2.68 \\
                              & $^{3} \mathrm{He}$ & 0.0512 & 0.0540 & 0.127 & 0.130 & 0.230 & 0.368 & 0.383 & 0.514 \\
                              & $\alpha$ & 12.5 & 11.5 & 18.2 & 18.0 & 20.3 & 24.4 & 24.7 & 28.6 \\
    \hline
    $^{40} \mathrm{Ca}$ 
                              & $p$ & 101 & 136 & 169 & 221 & 166 & 203 & 202 & 238 \\
                              & $d$ & 2.28 & 2.62 & 5.01 & 5.58 & 5.42 & 5.87 & 7.74 & 8.17 \\
                              & $t$ & 0.286 & 0.331 & 0.733 & 0.808 & 0.934 & 1.07 & 1.46 & 1.58 \\
                              & $^{3} \mathrm{He}$ & 0.0948 & 0.110 & 0.250 & 0.276 & 0.339 & 0.397 & 0.548 & 0.630 \\
                              & $\alpha$ & 25.7 & 25.5 & 39.3 & 42.4 & 38.5 & 39.6 & 42.0 & 43.0 \\
    \hline \hline
  \end{tabular}
\end{table*}
\clearpage
%

%
%